\documentclass[12pt]{iopart}
\usepackage{iopams}
\expandafter\let\csname equation*\endcsname\relax
\expandafter\let\csname endequation*\endcsname\relax
\usepackage{amsmath}

\begin{document}
\title{Note on Jackson's formalism of gauge transformation}
\author{V. Hnizdo} 
\address{2044 Georgian Lane, Morgantown, WV 26508, USA}
\eads{\mailto{hnizdo2044@gmail.com} }
\begin{abstract}
An outline is given of how Jackson may have obtained the  inhomogeneous wave equations for the auxilliary functions $\Psi$ and $\bf V$ in his  influential 2002 AJP paper on the transformation from the Lorenz gauge to other electromagnetic gauges. It clarifies the roles of these functions in the calculation of 
the Coulomb-gauge vector potential ${\bf A}_C$ by showing that while ${\bf A}_C$ is given directly  by 
$\bnabla\times \bf V$, only the  subtraction of $\bnabla \Psi$ from the Lorenz-gauge vector potential 
${\bf A}_L$ yields ${\bf A}_C$.
\end{abstract}
\section*{}
In his influential 2002 Am.\,J.\,Phys.\,paper on the transformation from the Lorenz gauge to other electromagnetic gauges [1], Jackson does not derive the inhomogeneous wave equations (2.10) satisfied by the auxiliary functions $\Psi$ and $\bf V$ of the formalism he develops for the transformation from the Lorenz gauge to the  Coulomb gauge.  To clarify the roles of these functions in the calculation of the Coulomb-gauge vector potential, we outline in this note how Jackson may have obtained his Eqs.\,(2.10). This will show that while  the vector potential in the Coulomb gauge is given directly by the curl of the function
$\bf V$, only  the subtraction of the gradient of the function  $\Psi$ from the vector potential in the Lorenz gauge yields the Coulomb-gauge vector potential.

The well-known inhomogeneous wave equation for the Lorenz-gauge vector potential ${\bf A}_L$
reads
\begin{align}
\Box\,{\bf A}_L =-(4\pi/c)\,{\bf J},
\label{AL}
\end{align} 
where ${\bf J}$ is the electric current density, which can be decomposed into its longitudinal and transverse components ${\bf J}_l$ and ${\bf J}_t$, respectively. Decomposing similarly also the vector potential
${\bf A}_L$, Eq.\,(\ref{AL}) will read  
\begin{align}
\Box\,({\bf A}_{L\,l}+{\bf A}_{L\,t}) =-(4\pi/c)\,({\bf J}_l+{\bf J}_t),
\label{BoxAL}
\end{align} 
where ${\bf A}_{L\,l}$ and ${\bf A}_{L\,t}$ are the longitudinal and transverse components of the Lorenz-gauge vector potential ${\bf A}_L$, respectively. Since the Coulomb-gauge vector potential ${\bf A}_C$,  which satisfies the inhomogeneous wave equation
\begin{align}
\Box\,{\bf A}_C= -(4\pi/c)\,{\bf J}_t,
\end{align}
is the transverse component of the vector potential in any gauge \cite{VH}, ${\bf A}_{L\,t}={\bf A}_C$  and so Eq.\,(\ref{BoxAL}) reduces to
\begin{align}
\Box\,{\bf A}_{L\,l} =-(4\pi/c)\,{\bf J}_l.
\end{align}
With ${\bf A}_{L\,l}$ written as $\bnabla \Psi$, using the fact that a longitudinal vector component can be expressed as the gradient of a scalar, and with ${\bf J}_l$ given by the expression for the longitudinal component of a current $\bf J$ in terms of the pertinent electric charge density $\rho$, 
Eq.\,(4) can be written as 
\begin{align}
\Box\,(\bnabla \Psi)=\bnabla\,(\Box \Psi)=-\frac{1}{c}\,\bnabla\int\frac{d^3 x'}{R}\frac{\partial \rho({\bf x}',t)}{\partial t},
\end{align}
where $R=|{\bf x}-{\bf x}'|$. Taking this equation to imply that
\begin{align}
\Box\,\Psi =-\frac{1}{c}\,\int\frac{d^3 x'}{R}\frac{\partial \rho({\bf x}',t)}{\partial t}
\label{BoxPsi}
\end{align}
yields the first of Jackson's equations (2.10). 

Expressing the transverse component ${\bf J}_t$ of $\bf J$  by
\begin{align}
{\bf J}_t= \frac{1}{4\pi}\,\bnabla\times\left(\bnabla \times\int \frac{d^3x'}{R}\,{\bf J}({\bf x}',t)\right),
\end{align}
and writing the transverse vector ${\bf A}_C$ as the curl of a vector, $\bnabla \times {\bf V}$, the inhomogeneous wave equation (3) for the Coulomb-gauge vector potential will read
 \begin{align}
\Box \,(\bnabla \times{\bf V})=\bnabla\times(\Box{\bf V}) =-\frac{1}{c}\,\bnabla\times\left(\bnabla \times\int \frac{d^3x'}{R}\,{\bf J}({\bf x}',t)\right).
\end{align}
Taking this equation to imply that 
\begin{align}
\Box \,{\bf V} = -\frac{1}{c}\,\bnabla \times\int \frac{d^3x'}{R}\,{\bf J}({\bf x}',t)
\label{BoxV}
\end{align}
yields the second of Jackson's equations (2.10).

Having decomposed the electric current density into longitudinal and transverse parts in Eq.\,(2.8), Jackson introduces  the auxiliary functions $\Psi$ and $\bf V$ by ${\bf A}_l=\bnabla\Psi$ and ${\bf A}_t=\bnabla\times{\bf V}$ in Eq.\,(2.9), where ${\bf A}_l$ and ${\bf A}_t$ are respectively  longitudinal 
and transverse parts into which, he says, ``a vector potential can be decomposed  in the same way as the current."  He then introduces the inhomogeneous equations (2.10) satisfied by the functions $\Psi$ and $\bf V$ by saying: ``{\it For the Coulomb-gauge vector potential} [my emphasis], the  auxiliary functions $\Psi$ and  $\bf V$ satisfy [these equations]." 
This is a confusing statement since it appears to imply that the Coulomb-gauge vector potential has both the longitudinal  and transverse  parts, $\bnabla\Psi$ and $\bnabla\times {\bf V}$. The Coulomb-gauge vector potential ${\bf A}_C$ is by definition a transverse vector, and as such is given fully by the transverse part 
$\bnabla \times {\bf V}$, which also equals the transverse part of the Lorenz-gauge vector potential ${\bf A}_L$. For the Coulomb-gauge vector potential ${\bf A}_C$, the function 
$\Psi$ plays a role only through the Lorenz-gauge vector potential ${\bf A}_L$ by 
${\bf A}_C={\bf A}_L-\bnabla\Psi$.

Deservedly, Jackson's paper  has been cited many times since its publication in 2002 -- still, to the present author's knowledge, nobody has pointed out yet the confusing  introduction there of the inhomogeneous wave functions (2.10) satisfied by the auxiliary functions $\Psi$ and $\bf V$. It may have confused some readers,  but it should not have  those who are well-versed in the subject, since equations (2.10) and the auxiliary functions $\Psi$ and $\bf V$ are used  in the paper in accordance with their true meaning.

The auxiliary function $\Psi$ plays an important role in Jackson's paper since the inhomogeneous wave equation (\ref{BoxPsi}) it satisfies  is, except for the different sign of its RHS, the same as 
his inhomogeneous equation (3.8) for the gauge function $\chi_C$ of the  transformation of the Lorenz-gauge potentials to those in the Coulomb gauge. The retarded solutions  of these inhomogeneous equations are thus the same except for their signs.

The function $\Psi$ can be calculated  for a  point charge $q$ moving  with constant velocity $v$ along the 
$z$-axis using its Lorenz-gauge vector potential,
\begin{align}
{\bf A}_L({\bf x},t)=\frac{\beta\gamma q}{[s^2+\gamma^2(z-vt)^2]^{1/2}}\,\hat{\bf z}, \quad \beta =v/c,\quad
\gamma=(1-\beta^2)^{-1/2},
\end{align}
where $s=(x^2+y^2)^{1/2}$.
The longitudinal part of ${\bf A}_L$ is given by
\begin{align}
{\bf A}_{L\,l}({\bf x},t)=-\frac{1}{4\pi}\bnabla \int\frac{d^3x'}{R}\,\bnabla'\bdot{\bf A}_L({\bf x}',t),
\end{align}
where the integral can be evaluated  by expanding the inverse distance $1/R$ in cylindrical coordinates (see \cite{Jackson}, p.\,140). 
Carrying out the evaluation of this integral  as in  a cylindrical-coordinate calculation of the gauge function $\chi_C$ for this case in \cite{HV}, we obtain
\begin{align} 
{\bf A}_{L\,l}({\bf x},t)=\frac{q}{\beta}\,\bnabla\left[{\rm arsinh}\frac{\gamma(z-vt)}{s}-{\rm arsinh}\frac{z-vt}{s}\right].
\end{align}
Since the function $\Psi$ is defined  by ${\bf A}_{L\,l}=\bnabla\Psi$, this gives $\Psi$ as
\begin{align}
\Psi({\bf x},t)=\frac{q}{\beta}\left[{\rm arsinh}\frac{\gamma(z-vt)}{s}-{\rm arsinh}\frac{z-vt}{s}\right].
\end{align}
Noting the expression for $\chi_C$ obtained in \cite{HV} and, by different means, in \cite{HV2}, this is in agreement with Jackson's result $\Psi=-\chi_C$.

We note that an alternative approach to the problem of the transformation of electromagnetic potentials in a given gauge to those in a different gauge has been developed in \cite{YMcD}.

\end{document}